\documentclass[aps,prd,showpacs,12pt]{revtex4-1}
\usepackage{graphicx}
\usepackage{bm}
\usepackage{pdfpages}
\input{epsf}
\begin{document}
\epsfverbosetrue
\def\la{{\langle}}
\def\ra{{\rangle}}
\def\vep{{\varepsilon}}
\newcommand{\beq}{\begin{equation}}
\newcommand{\eeq}{\end{equation}}
\newcommand{\beqa}{\begin{eqnarray}}
\newcommand{\eeqa}{\end{eqnarray}}
\newcommand{\q}{\quad}
\newcommand{\h}{\hat{H}}
\newcommand{\ha}{\hat{h}}
\newcommand{\p}{\partial}
\newcommand{\A}{|\Omega'|}
\newcommand{\AC}{{\it AC}}
\newcommand{\n}{\\ \nonumber}
\newcommand{\om}{\omega}
\newcommand{\Om}{\Omega}
\newcommand{\os}[1]{#1_{\hbox{\scriptsize {osc}}}}
\newcommand{\cn}[1]{#1_{\hbox{\scriptsize{con}}}}
\newcommand{\sy}[1]{#1_{\hbox{\scriptsize{sys}}}}
%\draft
\title{The meaning of "anomalous weak values" in quantum and classical theories}
%\title{No 'anomalous' weak values in a classical theory}
%\author {D. Sokolovski$^{a,b}$}
%\affiliation{$^b$ IIKERBASQUE, Basque Foundation for Science, Maria Diaz de Haro 3, 48013, Bilbao, Spain}
%\affiliation{$^c$ Departmento de F\' isica Aplicada I, EUITMOP, Universidad del Pa\' is Vasco, UPV-EHU, Bilbao, Spain}
%\affiliation{$^d$ Department of Physics, Shanghai University, 200444 Shanghai, China}
%\date{\today}
\begin{abstract}
%'Classical weak values' of Ferrie and Combes are by  no means 'anomalous'
\end{abstract}
%\title{Why there are no  'anomalous' weak values in a classical theory}
%\title{No  'anomalous' weak values in a classical theory}
\author {D. Sokolovski$^{a,b}$}
%\author {L:M. Baskim$^c$}
%\author {J. G. Muga$^{a,d}$}
\affiliation{$^a$ Departamento de Qu\'imica-F\'isica, Universidad del Pa\' is Vasco, UPV/EHU, Leioa, Spain}
\affiliation{$^b$ IKERBASQUE, Basque Foundation for Science, Maria Diaz de Haro 3, 48013, Bilbao, Spain}
%\affiliation{$^c$ Departmento de F\' isica Aplicada I, EUITMOP, Universidad del Pa\' is Vasco, UPV-EHU, Bilbao, Spain}
%\affiliation{$^d$ Department of Physics, Shanghai University, 200444 Shanghai, China}
%\date{\today}

\begin{abstract}
%The authors of  a recent paper [Phys. Rev. Lett. \textbf{113}, 120404 (2014)] suggest that "weak values are not inherently quantum but rather a purely statistical feature of 
%pre- and postselection with disturbance". We argue that this claim is erroneous,
%since such values require averaging with distributions which change sign. 
%This type of averaging  arises naturally in quantum mechanics, 
%but may not occur in classical statistics.
The readings of a highly inaccurate "weak" quantum meter, employed to determine the value 
of a dichotomous variable $S$ without destroying the interference between the alternatives,
may take arbitrary values.
We show that the {\it expected values} of its readings may take any real value, depending 
on the the choice of the states in which the system is pre- and post-selected.
Some of these values must fall outside the range 
 of eigenvalues of $A$, in which case they may be expressed as "anomalous" averages obtained with negative
 probability weights, constructed from available probability amplitudes.
 This behaviour is a natural consequence of the Uncertainty Principle. 
 The phenomenon of "anomalous weak values" has no non-trivial analogue in classical statistics.

\end{abstract}
%\date{\today}
\maketitle
\vskip0.5cm
            % pls. do not remove this line

%
%
%
\section {Introduction}
%The authors of \cite{PRL} suggest that "the phenomenon of anomalous weak values is not limited to quantum theory". 
%We argue that anomalous values cannot occur in a classical theory, and give our reasons below.
The interest in the so-called 'weak measurements' began with the publication of Ref.\cite{AHAR} entitled "How the result of a measurement of a component of the spin of a spin-1/2 particle can turn out to be 100". Recently the argument was extended to purely classical domain in \cite{PRL}, where the authors set to explain  "How can a result of a single coin toss turn out to be 100 heads".
The questions of this type invite two possible answers: either the measurement is not particularly good, or an error has been made in the analysis. Below we will show that the former is true in the case of \cite{AHAR}, and the latter - in the case of Ref. \cite{PRL}.
A hint of what happens in the quantum case can be taken from D. Bohm's warning \cite{BOHM} that 
"if the interference were not destroyed", "the quantum theory could be shown to lead to absurd results".
%Quantum weak measurements can and have been performed \cite{W1}-\cite{W2}, but the origin and physical meaning of quantum weak values remains open to further discussion.  
The claim that a simple
classical model may exhibit non-trivial anomalous weak values \cite{PRL} is, on the other hand, based on a simple misunderstanding. 
%
%One needn't read much beyond the title of \cite{PRL} to see that the answer to its question must be 'It cannot'. No matter how elaborate the protocol, some Alice at the end of the line wold receive the coin and write down the value of $1$ or $-1$, depending on whether the coin shows up heads or tails . Adding these numbers, and dividing them by the number of trials $N$, would always yield a value between $-1$ and $1$. So what is wrong with the argument of \cite{PRL}?
\newline
Other critique of Ref.\cite{PRL} can be found in \cite{COMM}, and we refer the interested reader to the Refs. in the Dressel's Comment
 \cite{COMM} for some further developments in the field of quantum weak measurements. 
\section{'normal' and 'anomalous' averages }
Consider an average of the form 
\begin{eqnarray}\label{1}
\bar{s}=\sum_{n=1}^N s_n P_n,\q \sum_n P_n=1
\end{eqnarray}
where $s_1>s_2...>s_N$. We will call $\bar{s}$ {\it normal} if it lies between $s_1$ and $s_N$, 
$s_1\ge \bar{s} \ge s_N$, and {\it anomalous} otherwise. It is readily seen that 
$\bar{s}$ is always normal if $P_n\ge 0$, $n=1,2,..N$, and to be anomalous it requires that 
at least one of the $P_n$ is negative. To see how an anomalous average may be produced, consider
$N=2$, $s_{1,2}=\pm1$, $P_1=1001$, $P_2=-1000$, $P_1+P_2=1$, and find that
$\bar{s}=2001$. Thus, a large anomalous value would occur where the moduli of $P_n$ are large, 
but the sum of all $P_n$ is unity due to a very precise cancellation.
Multiplying the $P_n$'s by $s_n$ destroys the cancellation, so that the resulting $\bar{s}$  is unduly large.
A more detailed example is given in \cite{ANN}, where a similar effect is found responsible for what appears
to be super-luminal transmission of a wave packet across a potential barrier. 
Next we discuss in detail the appearance of such alternating distributions in quantum measurement theory. 
\newline
%Anomalous averages arise naturally in 
\section{Quantum measurements with post-selection}
%The place to look is in quantum mechanics.
Consider a two level quantum system (a spin $1/2$)  with a hamiltonian $H$, pre- and post-selected (observed) in some states 
$\psi$ and $\phi$ at $t=0$ and $t=T$, respectively. Choose an operator $S$ with eigenstates $|s_{1,2}\ra$ and eigenvalues
$s_{1,2}\pm1$, e.g., the $z$-component of the spin, $S=\sigma_z$. Inserting, at some $0<t'<T$ the unity $I=|s_{1}\ra\la s_1|+|s_{2}\ra\la s_2|$ into the transition amplitude
$\la \phi|\exp(-iHT)|\psi\ra$,  shows that the spin can reach the final state via two virtual 
routes, $\psi\to s_1 \to \phi$  and $\psi \to s_2 \to \phi$, as shown in Fig.1. Putting for simplicity $H=0$, for the two corresponding amplitudes we 
have
\begin{eqnarray}\label{1a}
A_{1,2}=\la \phi |s_{1,2}\ra\la s_{1,2}|\psi\ra.
\end{eqnarray}
\begin{figure}
	\centering
		\includegraphics[width=6cm,height=3cm]{{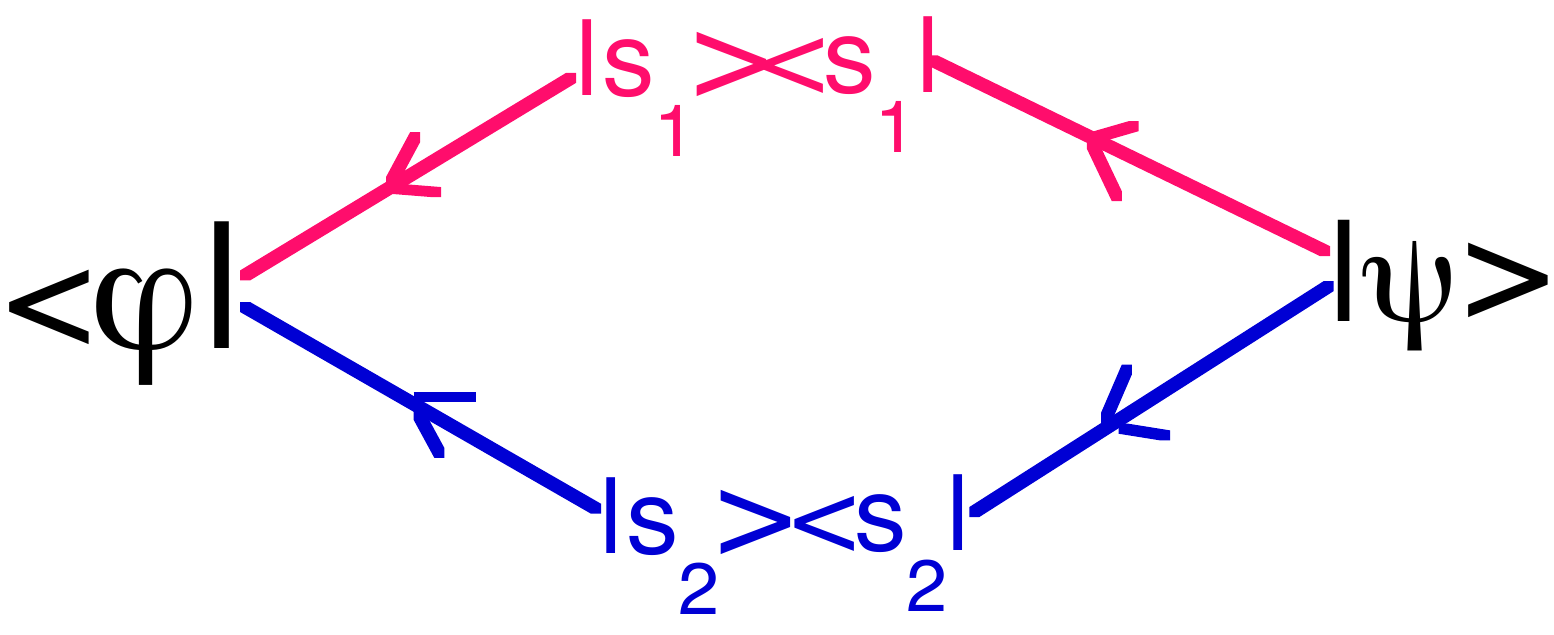}}
\caption{Two virtual routes connecting initial and final states $|\psi\ra$ and $|\phi\ra$.
A meter acts to destroy the interference between them.}
% $\alpha=c\tau$. For $c\tau >> 1$ note  
% $t\sim the pattern related to opening 
%channels at $ct=c\tau$. }
\label{fig:4}
\end{figure}
To see what actually happens at $t'$
% we may couple to the spin 
%We 
we may employ a von Neumann pointer with the position $f$, 
%and the momentum $p=-i\partial_f$, 
%prepared 
%at $t=0$ in 
initially 
 decoupled from the spin.
 We set the pointer at some $f'$ by preparing it in a  state $|M_{f'}\ra=\int df G(f-f') |f\ra$, where $G(f)$ is a function 
 peaked around $0$ with a width $\Delta f$, such that $\int|G(f)|^2=1$.
  For  $t' \le t''\le t'+\tau<T$ the pointer briefly interacts with the spin via
$H_{int}=-i\tau^{-1}\partial_f S$, and then its final position is measured (read) exactly. 
\newline
We can do the measurement in three steps \cite{FOOT0}. First the meter acts, after which the entangled state of the system becomes 
\begin{eqnarray}\label{1ae}
%\la f|\Phi\ra=G(f-s_1)\la s_1|\psi\ra |s_1\ra+G(f-s_2)\la s_2|\psi\ra |s_2\ra. 
|\Phi\ra= \q\q\q\q\q\q\q\q\q\q\q\n
\int [G(f-s_1)\la s_1|\psi\ra |s_1\ra+G(f-s_2)\la s_2|\psi\ra |s_2\ra]|f\ra df.
\end{eqnarray}
Then the pointer's reading is found to  be $f$, which leaves the spin polarised in along some axis in an (unnormalised) pure state  $\la f|\Phi\ra$. Finally, in this state we measure the projector on $|\phi\ra$, and keep the results only if the projection is successful, which happens with a probability $|\la \phi|\la f|\Phi\ra|^2$. For the (unnormalised) probability of a reading $f$ in our pre- and post-selected setup we have
\begin{eqnarray}\label{2a}
P(f|\phi\gets \psi,\Delta f)=|B_1(f)+B_2(f)|^2=\n
|B_1(f)|^2+|B_1(f)|^2+2Re[B_1(f)B_2^*(f)],
\end{eqnarray}
where $B_{1,2}=G(f-s_{1,2})A_{1,2}$.
Choosing, with no loss of generality, a Gaussian pointer, 
\begin{eqnarray}\label{1ae}
 G(f)=(2/\pi \Delta f^2)^{1/4}\exp(-f^2/\Delta f^2),
\end{eqnarray}
for the expected value of the pointer's reading, $\overline{f}$ we find
\begin{eqnarray}\label{3ba}
\overline{f}\equiv \int f P(f|\phi\gets \psi,\Delta f)df =
%\q\q\q
%\q\q\q\q\q\q\q\q
%\q\q\q\q\q\q\q
%\q\q
%\q\q\q\q
%\q\q\q\q\q\q\q\q\q\q\q\q\q\q
%\overline{f}=
%\n
% \frac{
 \{s_1|A_1|^2+s_2|A_2|^2\q\q\n
 +Re[A_1A_2^*](s_1+s_2)\exp[-(s_1-s_2)^2/2\Delta f^2]\}/N\q\q\n
 %}
N\equiv{|A_1|^2+|A_2|^2+2Re[A_1A_2^*]\exp[-(s_1-s_2)^2/2\Delta f^2]}.\q
%\end{small}
\end{eqnarray} 
Equation (\ref{3ba}), expresses $\overline{f}$ in terms of the parameters which describe the measured variable and the transition, $A_{1,2}$ and $s_{1,2}$. It does not have the form (\ref{1}), except it two special cases. We consider these cases next.
\section{Accurate "strong" measureaments}
%, and its width $\Delta f $ is small,
 Now $\Delta f $ determines what is known about the initial position of the pointer and, therefore, the accuracy of the measurement. 
 By sending $\Delta f\to 0$ we make the pointer position correlate exactly with the eigenvalues of $S$, $s_{1,2}$, so that
 \begin{eqnarray}\label{2am}
P(f|\phi\gets \psi,\Delta f\to 0)=\n
|A_1|^2\delta(f-s_1)+|A_2|^2\delta(f-s_2).\q\q
\end{eqnarray} 
With no overlap between $G(f-s_1)$ and $G(f-s_2)$, finding a reading $f=s_i$ leaves the spin in the state $|s_i\ra$, $i=1,2$.
The average meter reading (\ref{3ba}) now has the form (\ref{1}), 
 \begin{eqnarray}\label{2a1}
\overline{f}=s_1P^{strong}_1+s_2P^{strong}_2\equiv \overline{s}_{strong},
\end{eqnarray} 
where
 \begin{eqnarray}\label{2a2}
P^{strong}_i\frac{|A_i|^2}{|A_1|^2+|A_2|^2},\q i=1,2
\end{eqnarray}
are non-negative probability weight. Thus, expressed in terms of $A_{1,2}$ and $s_{1,2}$, $\overline{s}_{strong}$ is always  a "normal" average. This is just a way of saying that, for all $\psi$ and $\phi$,  the average reading of an accurate meter always lies between $s_1$ and $s_2$.
%%%%%%%%%%%%%%%%%%%%%%%%%%%%%%%%%%%%
\section{A 'classically' inaccurate meter}
Suppose next that we still have an accurate 'strong' meter with $\Delta f << s_1-s_2$, but for some reason are unable to
set it precisely to zero. Rather, initially the pointer reads $f'$ with a probability $W(f')=W(-f')$, 
in a range of a width $\delta f'$ around zero.
The initial state is now mixed,
\begin{eqnarray}\label{3}
\rho_M=\int|M_{f'}\ra W(f')\la|M_{f'}|df',   \q \int W(f')df'=1,\q
\end{eqnarray}
and finding a final reading $f$ leaves the spin in a mixed state 
\begin{eqnarray}\label{3m}
\rho_{spin}^{final}=|s_1\ra W(f-s_1)\la s_1|+|s_2\ra W(f-s_2)\la s_2|. 
\end{eqnarray}
For $\delta f'>>s_1-s_2$, the pointer's final readings are distributed with the probability 
$P(f|\phi\gets \psi,\Delta f|\delta f')=\int P(f-f'|\phi\gets \psi,\Delta f)W(f') df'$ over a broad range $\sim \delta f'$. The meter appears to have lost correlation with the eigenvalues $s_{1,2}$.
Yet some information about them may be recovered, provided one is  only interested in average values. Thus, for $\delta f' >> s_1-s_2$ and $\Delta f << s_1-s_2$, the average pointer position is still given by Eq. (\ref{2a1}) \cite{ME}
\begin{eqnarray}\label{3a}
\overline{f}= \int f P(f|\phi\gets \psi,\Delta f\to 0|\delta f')df 
=
%s_1P^{strong}_1+s_2 P^{strong}_2,
%\equiv  
\bar{s}_{strong}.
%\q P^{strong}_{1,2}= |A_{1,2}|^2/(|A_{1}|^2+|A_{2}|^2)
%\n  \frac{s_1|A_{1}|^2+s_2|A_{2}|^2}{|A_{1}|^2+|A_{2}|^2}\equiv \bar{s}_{strong}.\q
\end{eqnarray}
No matter how large the {\it classical} uncertainty, the mean pointer reading would still lie between $s_1$ and $s_2$.
A different result is achieved if the initial pointer position is made uncertain in the {\it quantum} sense.
%%%%%%%%%%%%%%%%%%%%%%%%%%%%%%%%%%%%%%%%
\section{How 'negative probabilities' enter quantum measurement theory}
Next consider a pointer, prepared initially in a very broad pure state,  $\Delta f \to \infty$. The accuracy of the measurement is very low, since the initial pointer position is highly uncertain, albeit in a different sense. There is only a probability amplitude, $G(f)$, and not the probability, for it to be set to a particular $f$. As in Sect. V, the probability distribution of pointer's reading is very broad, so that $\overline{f}$ can, in principle, take any  values as $\Delta f\to \infty$.

With  both exponentials in Eq.(\ref{3ba}) tending to unity, for $\overline{f}$ we have
\begin{eqnarray}\label{3d} 
\overline{f}=s_1P^{weak}_1+s_2 P^{weak}_2\equiv Re \bar{s}_{weak},
\end{eqnarray}
where
\begin{eqnarray}\label{3c}
%\overline{f}\equiv \int f P(f|\phi\gets \psi,\Delta f|\delta f')df =
P^{weak}_{1,2}=Re\frac {A_{1,2}}{A_1+A_2},
\end{eqnarray}
%can be of either sign and 
$ \bar{s}_{weak}=(s_1A_1+s_2A_2)/(A_1+A_2)$ is quantum mechanical {\it weak value} of the operator $S$, which can take complex values (for a recent discussion see, for example \cite{JAPS}).
% for the transition $\psi \to \phi$. 
For $P^{weak}_{1,2}$ having opposite opposite signs, 
%we can say that, expressed in terms of $A_{1,2}$ and $s_{1,2}$,
$Re \overline{s}_{weak}$ in Eq.(\ref{3d}) has the appearance of an "anomalous" average obtained with negative probability weights \cite{FOOT1}.
This is an elaborate way to say  that  the average reading of a meter, highly inaccurate in the quantum 
sense,  is not {\it apriori} restricted in its magnitude.
 %may lie almost anywhere at all, depending on the choice of the transition $\phi\gets \psi$ \cite{FOOT1}.  
%%%%%%%%%%%%%%%%%%%%%%%%%%%%%%%%%%%%%%%%%%%%%
\section{Anomalous values and the Uncertainty Principle}
One  purpose of this paper is to establish why the meters in Sect. V and Sect. VI behave so differently.
%It is helpful to re-examine the question in slightly different terms. 
Suppose that one chooses a transition, 
and after series of weak measurements finds a mean value of $100$ for a spin of 1/2. We may suspect that 
this is a 'wrong' result obtained with a malfunctioning meter. The problem is, we cannot produce the 'correct' answer,
and a brief look into quantum mechanical text book shows that it may not even exist.
 \newline
Equation (\ref{2a}) suggests that we are dealing with a simple version of Young's two-slit experiment. The pointer "arrives' to a 'point on the screen" $f$ by passing through "two slits", corresponding to spin values of $1$ and $-1$, with the probability amplitudes $B_1(f)$ and $B_2(f)$. 
%Thus, by measuring $S$ we are, in effect, trying to answer the 'Which way?' question. 
The problem is well known in literature \cite{Feyn}. A strong measurement of Sects. IV and V destroys the interference between the paths, 
converting two virtual routes into two real ones, to each of which one can now ascribe a probability.
A weak measurement of Sect.VI leaves the interference intact, and the probabilities in Eq.(\ref{2a}) contain 
an interference term, which involves both virtual routes at the same time (cf. Fig.2). 
\begin{figure}
	\centering
		\includegraphics[width=6cm,height=6cm]{{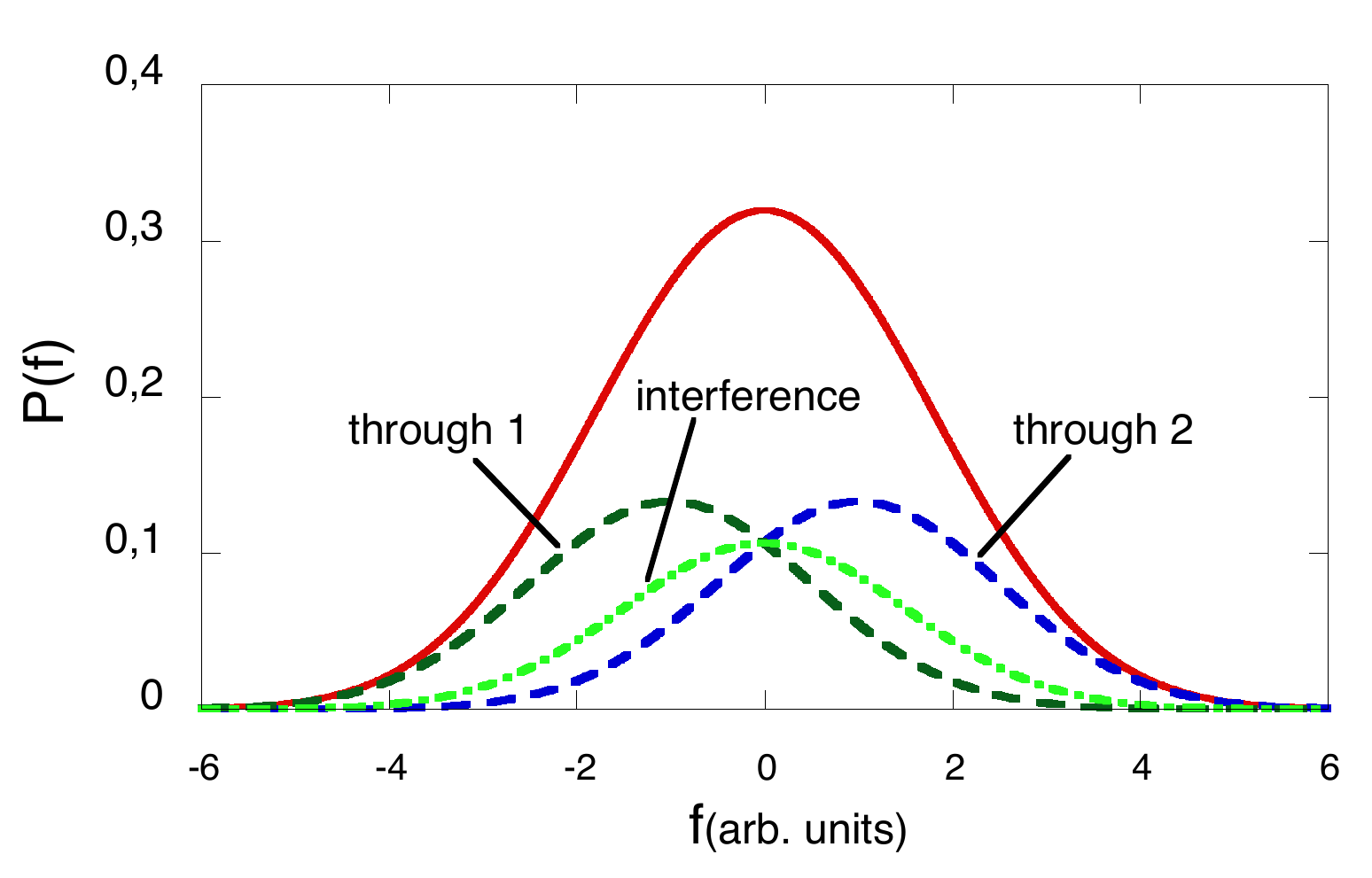}}
\caption{Three contributions to the probability $P(f)$ in Eq.(\ref{2a}) (solid) for $|\psi\ra=|\phi\ra=(|s_1\ra+|s_2\ra)\sqrt{2}$ and $\Delta f=3$.}
% $\alpha=c\tau$. For $c\tau >> 1$ note  
% $t\sim the pattern related to opening 
%channels at $ct=c\tau$. }
\label{fig:4}
\end{figure}
One's inability to refer the interference term to any one route is reflected in the Uncertainty Principle, which states that two interfering routes cannot be told apart and should be considered a single pathway \cite{Feyn}.
\newline
In our example, different values of the spin's components label different virtual paths in Fig. 1, and the mean value of $S$ may tell us something about how these paths are travelled. Obtaining in a series of strong measurements $\bar{s}_{strong}>0$ allows to conclude
that the lower route in Fig.1 is travelled more frequently, and vice versa. No such conclusion can be drawn if the measurements are weak. If by asking
%If so, the question 
{\it "what was, on average, the value of $S$ if we hadn't destroyed coherence between the two routes?"} 
one hoped to learn something about which path was actually travelled, he must be disappointed.  The bizarre mean value 
of $100$ stands as a reminder that indivisible cannot, after all, be divided \cite{FOOT}.
\newline
%should not have a meaningful answer \cite{ME},. Although a mean value of $100$ for a spin of 1/2 looks like a meaningless result produced by a malfunctioning meter, it cannot simply be "corrected", since the "correct" answer just doesn't exist.
Still, this result needs to be interpreted, and there is a choice: to follow Bohm \cite{BOHM} in discarding it as "absurd", or to follow the authors of \cite{AHAR} in trying to ascribe to it a degree of importance and "reality", simply because such a measurement can be made? In the next Section we will 
extend Bohm's argument in favour of strong measurements, and look at all, rather than just one, possible transitions.
% which out two-level system may undergo. 
%In other words, the question is about the physical meaning of quantum weak values, and we will address it in the next Section.
\section{An answer to the awkward question}
\begin{figure}
	\centering
		\includegraphics[width=8.5cm,height=5cm]{{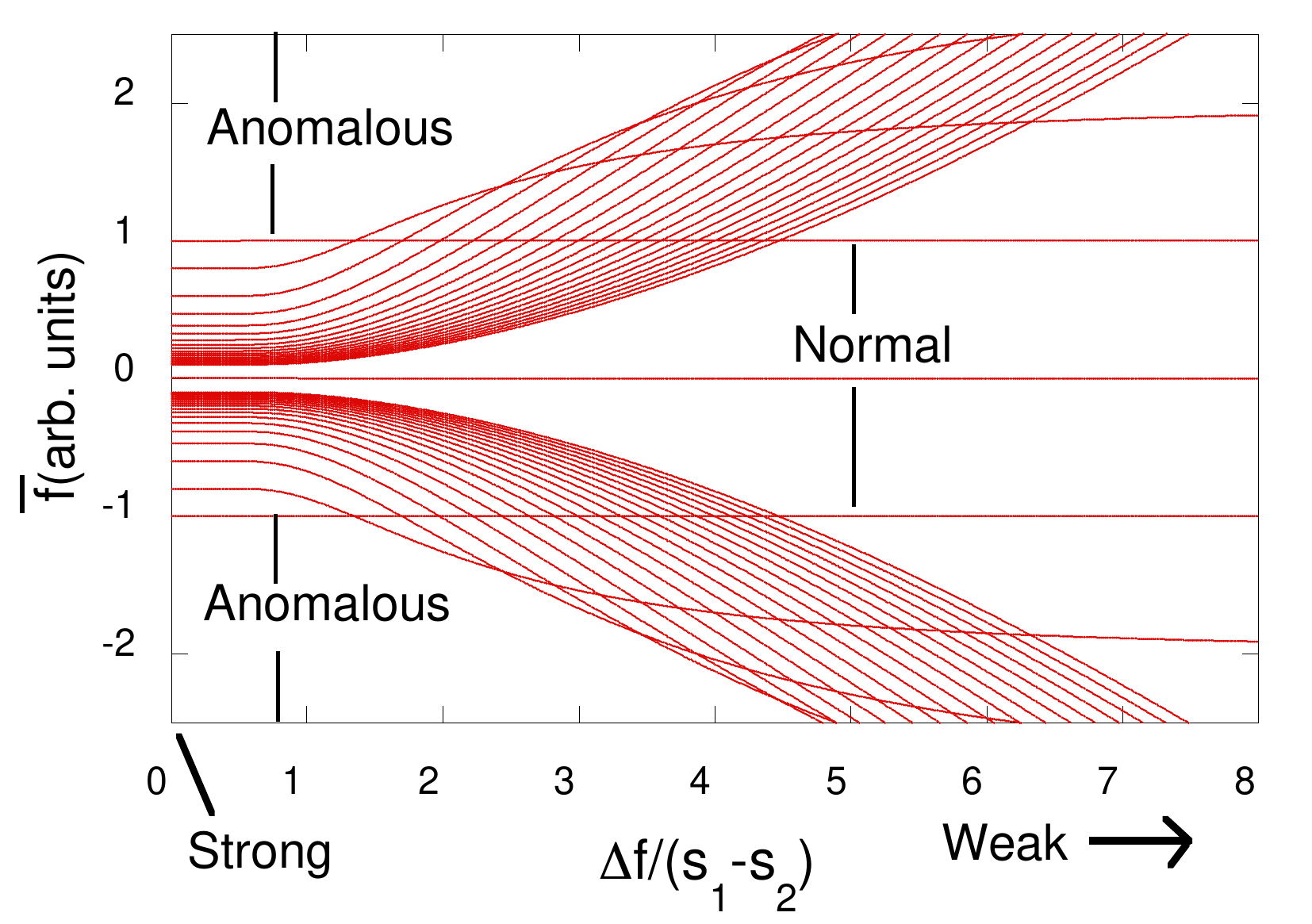}}
\caption{Dependence of the mean meter reading $\overline{f}$ on the accuracy of the measurement, $\Delta f$ for
$|\psi\ra=[|1\ra+|2\ra ]/\sqrt{2}$. Final states $|\phi\ra$ are chosen so that, as $\Delta f \to \infty$, $\overline{s}_{weak}=-20, -21...0,...20$. }
% $\alpha=c\tau$. For $c\tau >> 1$ note  
% $t\sim the pattern related to opening 
%channels at $ct=c\tau$. }
\label{fig:a}
\end{figure}

Usually it is accurate  'strong' measurements, which provide one with useful information.
To know whether two or three holes have been actually cut in the screen, we shine on it a light 
of wave-length short enough (a strong measurement) to produce just two, and not three, bright spots behind it. Shining light of a very large wave length (a weak measurement) would produce a very broad spot, whose centroid [cf. Eq.(\ref{3a})] may lie far away from the two bright spots observed previously. Based on this, it would be hard to guess the number of holes actually made. 
\newline
Similarly, to find out what sort of spin has an electron, one can devise a Stern-Gerlach experiment \cite{BOHM} with pre- and post-selection for the spin variable in the states $|\psi\ra$ and $|\phi\ra$,
 \begin{eqnarray}\label{2aa}
|\psi\ra=\frac{|1\ra+a|2\ra}{\sqrt{1+|a|^2}},\q |\phi\ra=\frac{|1\ra+b|2\ra}{\sqrt{1+|b|^2}}.\q\q
\end{eqnarray}
%(whatever they might be).
 For {\it all} choices of these states an accurate measurement of $\sigma_z$  would produce only the values of $\pm1$. In all cases the average of $\sigma_z$ will lie between $-1$ and $1$. This is how one knows that spin of $1/2$ is an intrinsic property of the electron, 
and is later able to  write its wave function as a Pauli spinor in situations much more general than the original Stern-Gerlach setup.
\newline
Much less can be learned about the electron if only inaccurate weak measurements are made. It is sufficient to consider real valued $a$ and $b$ in equation (\ref{2aa}) to show that 
 \begin{eqnarray}\label{2ab}
\overline{f}\equiv Re \overline{s}_{weak}=(A_1-A_2)/(A_1+A_2)
\end{eqnarray}
can have any real value $Z$, $-\infty \le Z \le \infty$, provided
 \begin{eqnarray}\label{2ac}
ab=(1-Z)/(1+Z),
\end{eqnarray}
[see Fig. \ref{fig:a}].
In particular, the choice $b=a$ yields a "normal"  $\overline s_{weak}=\la \psi| A |\psi\ra$ which coincides with the expectation 
value of $A$ in the state $|\psi\ra$.
The choice $b=1/a$ leads to $A_1=A_2$, so that $\overline s_{weak}$ coincides with $\overline s_{strong}$, and also is "normal".
Finally, for $b=-99/101a$ $\psi$ and $\phi$ are nearly orthogonal, and $\overline s_{weak}=100$ is an "anomalous"  average first obtained in \cite{AHAR}.
Such values are properties of particular transitions, and tell very little about the electron's own properties \cite{FOOT3}.
The answer to the question  which, as we argued above should not have a meaningful answer, is, in this case,  {\it "anything at all, depending on the circumstances"}.
%%%%%%%%%%%%%%%%%%%%%%%
\section{A cat and its smile}
Let us illustrate the above conclusion by a more recent example \cite{CAT}. Consider a system consisting of two 2-level systems, 
pre- and  post-selected, as before,  in yet unknown states $\psi$ and $\phi$. There altogether four orthogonal projectors, two for each degree of freedom: $\Pi_L=|L\ra \la L|$, $\Pi_R=|R\ra \la R|$, $s_+=|+\ra \la+|$, $s_-=|-\ra \la-|$. Following \cite{CAT} we will say that 
the system is on the left (right) if it is in the eigenstate of $\Pi_{L(R)}$, and  its spin is up (down) along the $z$-axis, if it is in an eigenstate of $s_{+(-)}$.
With four possible intermediate measurements, there are four routes connecting $|\psi\ra$ with $|\phi\ra$, with the amplitudes
$A_{ij}=\la \phi|\Pi_is_j|\phi\ra$, $i=L,R$, $j=\pm$ \cite{RAFA}. 
There are also two spin operators, $\sigma^L_z=\Pi_L\sigma_z$ and $\sigma^R_z=\Pi_R\sigma_z$, $\sigma_z=s_+-s_-$, 
for a system found on the left, and on the right respectively. With both operators measured accurately, the four routes are travelled with the probabilities $P^{strong}_{ij}=|A_{ij}|^2$ and, obviously, the found value of $\Pi_L\sigma_z$ is zero, if the system has chosen the right route, and vice versa. The authors of \cite{CAT} have demonstrated that this is no longer the case, if the measurements are weak, so that the interference between the routes is not destroyed.  In particular, with the special choice of  $\psi$ and $\phi$ it is possible to have 
$\overline{\Pi_L}_{weak}=(A_{L+}+A_{L-})/\sum_{ij}A_{ij}=1$, $\overline{\Pi_R}_{weak}= (A_{R+}+A_{R-})/\sum_{ij}A_{ij} =0$,  $\overline{\sigma^L_z}_{weak}= (A_{L+}-A_{L-})/\sum_{ij}A_{ij} =1$, and $\overline{\sigma^R_z}_{weak}= (A_{R+}-A_{R-})/\sum_{ij}A_{ij} =1$. So the authors of \cite{CAT} conclude that the system (the 'cat') is in one place, while its spin (its 'smile') is elsewhere.
  
  Our interest here is to show that if we go over {\it all} possible transitions we will be able to find systems with {\it all} possible distributions of the spin between where the system is, and where it is not. To show this, we write 
 \begin{eqnarray}\label{2adt}
|\psi\ra =[\alpha_1|L\ra|+\ra+\alpha_2|L\ra|-\ra+\alpha_3|R\ra|+\ra+\alpha_4|R\ra|-\ra]/[\sum_{i=1}^4|\alpha_i|^2]^{1/2}, \n
|\phi\ra =[\beta_1|L\ra|+\ra+\beta_2|L\ra|-\ra+\beta_3|R\ra|+\ra+\beta_4|R\ra|-\ra]/[\sum_{i=1}^4|\beta_i|^2]^{1/2},
\end{eqnarray}  
and require that while the cat is weakly on the left side, its smile is distributed between left and right,
 \begin{eqnarray}\label{2adtt}
\overline{\Pi_L}_{weak}=1,\q \overline{\Pi_L}_{weak}=0,\q\overline{\sigma^L_z}_{weak}=X, \q \overline{\sigma^R_z}_{weak}=Y, 
\end{eqnarray} 
where $X$ and $Y$ are any real numbers. A simple algebra shows that for Eqs.(\ref{2adtt}) to hold, $\alpha$'s and $\beta$'s in Eq.(\ref{2adt}) may be chosen, for example, as
 \begin{eqnarray}\label{2adttt}
\alpha_1=\beta_1=1,\q \alpha_2=\beta^*_2=\sqrt{(1-X)/(1+X)},\q\n
\alpha_3=\beta^*_3=\sqrt{Y/(1+X)},\q \alpha_4=\alpha_3=-\beta^*_4.\q\q\q
\end{eqnarray} 
As in the previous Section, this choice is not unique, and there are enough parameters in the wave functions $\psi$ and $\phi$, to make any choice of the weak values (\ref{2adtt}) possible.  
Thus Alice, having chosen $\alpha_2=\beta_2=\alpha_3=\beta_3=0$, would find the smile firmly on the cat's face, $X=1$, $Y=0$.
Bob, working with $\alpha_2=\beta_2=\alpha_3=\beta_3=1$ finds, as did the authors of \cite{CAT}, the smile completely detached from the cat, $X=0$, and $Y=1$. Finally Carol, choosing, $\alpha_2=\beta_2=0$ and $\alpha_3=\beta^*_3=100i$, finds a smile where the cat is, $X=1$, but also a substantial "frown", $Y=-200$, where the cat isn't. Together, Alice, Bob, Carol and co-workers must conclude that  the only thing they learnt  about a cat's relation with its smile by making weak measurements, is that this relation may take any form at all.
Bob and Carol may be surprised by their results, and it is the prime objective of our discussion to demonstrate that they shouldn't be.

\section{The ways to make a measurement "weak"}
A brief remark is in order. Consider again the von Neumann Hamiltonian, this time with a adjustable parameter $\lambda$, 
 \begin{eqnarray}\label{2ad}
H_{int}=-i\lambda \frac{d}{df} S.
\end{eqnarray}
There are three equivalent ways to ensure that the meter perturbs the system as little as possible. 

(A) Reduce the interaction strength, $\lambda \to 0$ and leave the initial meter state $G(f)$  as it was , as was done in \cite{AHAR}. Hence the adjective "weak" widely used in this context.

(B) Put the coupling strength to unity, and rescale the meter's position $f\to f/\lambda$.
As $\lambda \to 0$ the width of the pointer's state increases, $\Delta f \to \Delta f/\lambda$.
The mean pointer position is then proportional to the weak value of the operator $S$, 
$\overline{f}\to \la\phi|S|\psi\ra/\la\phi|\psi\ra$. The value is "anomalous" should it lie outside the interval $[s_1,s_2]$. This is the convention we followed throughout this paper.

(C) 
%Leave the pointer's initial state as it was, and 
Incorporate $\lambda$ into the new operator $S'$, $S'=\lambda S$ with the eigenvalues $\lambda s_{1,2}$ and send $\lambda \to 0$.
The mean pointer position is then proportional to the weak value of the measured operator $S'$, 
$\overline{f}\to \lambda \la\phi|S'|\psi\ra/\la\phi|\psi\ra$. The value is "anomalous" should it lie outside the narrow interval $[\lambda s_1,\lambda s_2]$. 
\newline
%In brief, the mean meter reading coincides with the weak value of an operator $\lambda S$ provided the width of the pointer's pure state greatly exceeds $\lambda s_1 -\lambda s_2$.
In all cases we achieve the essential condition for the "weakness": the spectrum of the measured operator "fits" under the the Gaussian $G(f)$, so the interference between the two pathways is not destroyed. It is important, however, to follow the chosen convention  consistently, as we  illustrate in the next Section.   
%%%%%%%%%%%%%%%%%%%%%%%%%%%%%%%%%%%%%%%
\section{There are no anomalous mean values in a classical theory}
% so all the weights in Eq.(\ref{1}) are non-negative, and all classical averages are, therefore, "normal".
For each measurement scheme, quantum mechanics does in the end produce
a classical statistical ensemble with non-negative probabilities, so all quantum averages are of the normal type.  
Yet, in Sect. V it was shown that 'negative probabilities' enter the theory if we try to rewrite the "normal" mean position of a quantum  pointer, which has lost correlation with the measured system, in terms of the variables describing the system and the transition. Quantum mechanics operates with probability amplitudes from which one can construct for Eq.(\ref{1}) weights of either sign. 
 The Uncertainty Principle requires some of the averages written is this way to be "anomalous".
 \newline
Classical statistics operates only with non-negative probabilities.  There is no analogue of the Uncertainty Principle, and no quantities, similar to probability amplitudes, from which to construct negative weights in Eq.(\ref{1}). Thus, no matter how elaborate a measurement scheme,
%{\it 
no averages in a purely classical theory can be anomalous.
% } 
%\newline For example
\newline As an illustration, consider a strong measurement of an operator $S$ using the meter (\ref{2ad}) with $\lambda=1$. 
With the interference destroyed, the meter is classical, its outcomes are $\pm1$, and the mean of its readings, $\overline{s}$,  lie in the interval
$[-1,1]$. As was shown in Sect.V, making the initial position of such a meter uncertain, does not lead to anomalous values.
There are, however, two trivial ways to obtain a mean outside this interval: (i) shift the origin for the pointer by $f_0$, $G(f)\to G(f-f_0)$;
(ii) recalibrate the meter, so that a value $\alpha s$ is recorded instead of $s$, by choosing in (\ref{2ad}) $\lambda=\alpha$.
Obviously, the new mean values, $\overline s'=\overline {s}-f_0$ and $\overline {s}''=\alpha \overline {s}$ should not be considered 
'anomalous means' of the operator $S$, but rather the 'normal' means of $S+f_0$ and $\alpha S$, respectively. 
%What teh hell is the difference from the waek case??????
\section{An example}
%\newline The above statement should suffice as a criticism of \cite{PRL} where it is claimed
%that an analysis "demonstrates a simple
%classical model which exhibits anomalous weak values".
We conclude by briefly reviewing the original proposal in Ref. \cite{PRL}. In the scheme illustrated in Fig.4, Alice sends Bob a coin heads up which, on arrival, may change side with a probability $\alpha\ge 0$. Bob records the result (1 for heads and -1 for tails) on a piece of paper, and then sends the coin back to Alice. On its way the coin may again change the side, this time with a probability  $1-\delta \ge 0$, and Alice tells Bob to keep his note only if the coin arrives to her heads up. The authors of \cite{PRL} claim that the mean recorded by Bob is $1/(1-\delta)$, so that for $\delta=0.99$ "outcome of the toss coin is 100 heads". This, however, is not possible, for the  simple reason that  adding the $1$'s and $-1$'s, and dividing the sum by the number of notes he has kept, Bob would always get a result $a_w$  between $-1$ and $1$.  Equivalently, 
$a_w$ is a normal average (\ref{1}) obtained with non-negative weights $P_1=(1-\alpha)\delta/[(1-\alpha)\delta+(1-\delta)\alpha]$ and $P_2=\alpha(1-\delta)/[(1-\alpha)\delta+(1-\delta)\alpha]$ and, therefore, must be contained between $-1$ and $1$.
The only way for Bob to obtain the average of $100$ is to resort to one of the possibilities outlined in the previous Section.
Indeed, the authors of \cite{PRL} "recalibrate" Bob, making him calculate the mean not of $s$, but of $s$ divided by a small parameter $\lambda$, $s'=s/\lambda$ [cf. Eq. (19) of \cite{PRL}]. It is easy to check that in this way  they obtain a perfectly normal average of a larger variable $S/\lambda$.
Thus, it may be concluded that a result of a single toss of a coin is $100$ provided the values in excess of $100$, say, $\pm 200$ are attributed to the two sides of the coin. The suggestion that  anomalous weak values are "a purely statistical feature of pre- and post-tselection with disturbance" is, however, wrong \cite{FOOT2}.
\begin{figure}
	\centering
		\includegraphics[width=5cm,height=3cm]{{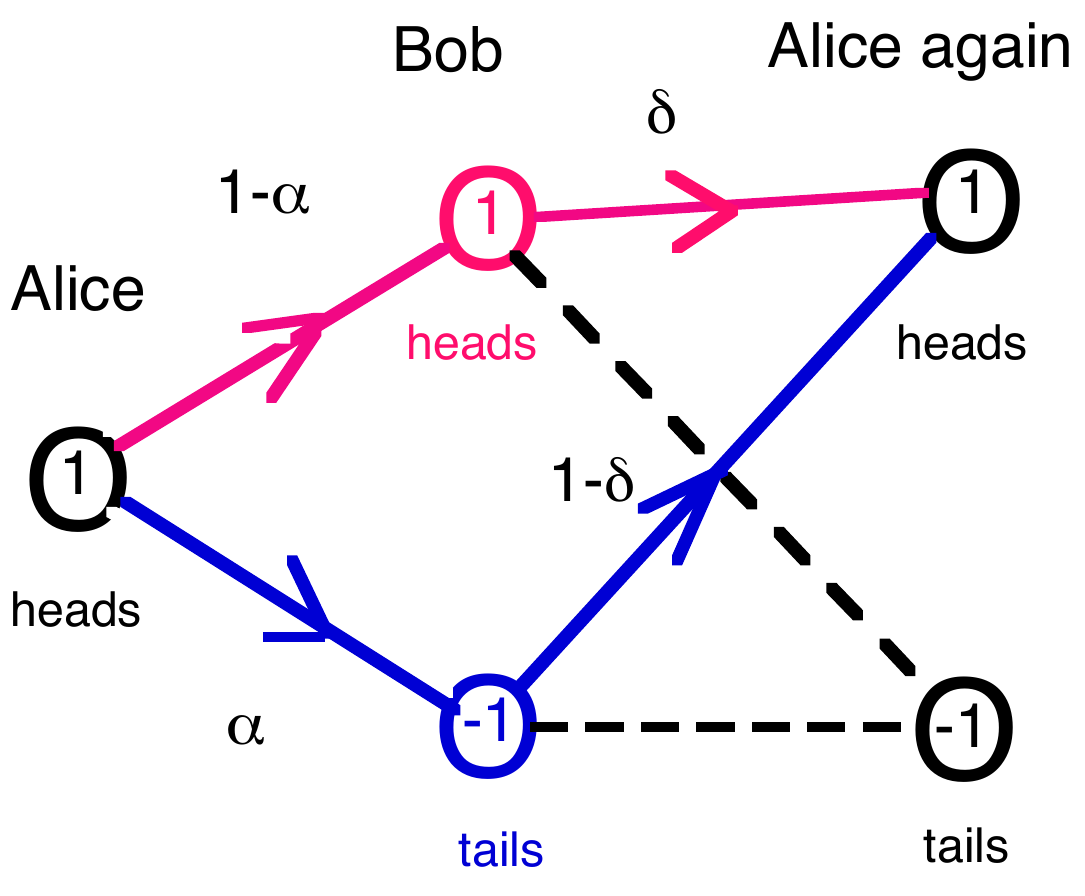}}
\caption{A coin, which may be flipped in transit, is passed from Alice to Bob and then returned to Alice, 
Bob keeps the statistics only if Alice receives it heads up.}
% $\alpha=c\tau$. For $c\tau >> 1$ note  
% $t\sim the pattern related to opening 
%channels at $ct=c\tau$. }
\label{fig:4}
\end{figure}
%%%%%%%%%%%%%%%%%%%%%%%%%%%%%%%%%%%%%%%
\section{Conclusions}
Some of the confusion (see, e.g.,\cite{PRL}) currently surrounding the issue of weak values, arises from the interpretation the authors of \cite{AHAR} have given to their original result.
It is, therefore, worth looking at the issue from a different prospective. A value of "100" obtained for a spin of $1/2$ appears "surprising" if one expects the observed values to be limited to the eigenvalue range of the measured operator $A$. In the weak limit, this expectation is, however, unfounded.
With its back action minimised, a meter is trying to distinguish between two routes combined into one by quantum interference.
Faced with a question which should have no answer, one has two possibilities.
One is to refuse to give an answer, a luxury not permitted to an experimental setup.
The other is to give an answer containing no information about the studied object, since the relevant information does not exist. Thus, for various transitions, a weak quantum measurement  produces results both 'normal' and 'absurd' in the sense of Bohm \cite{BOHM}, in fact, any results at all.
The anomalous expectation values are unique to quantum mechanics, and cannot arise in a purely classical theory, except in the trivial sense, as explained in Sect. X.
 \section{Acknowledgements:}
We acknowledge support of the Basque Government (Grant No. IT-472-10), and the Ministry of Science and Innovation of Spain (Grant No. FIS2009-12773-C02-01).


\begin{thebibliography}{10}
\bibitem{FOOT0} Often one chooses to first post-select the spin in its final state, and then read the meter.
Since the two operations commute, their order is of no importance. Our choice serves to emphasise the fact that an 
inaccurate meter leaves the spin in a rather complicated linear combination of the up- and down-states.
\bibitem{FOOT1}It is worth noting that the 'improper' distribution (\ref{3c}) doesn't fully replace $P^{strong}_{1,2}$ as $\Delta f\to 0$. While for a accurate strong measurement we have $\overline{f^2}=s_1^2P^{strong}_1+s_2^2P^{strong}_2$, $\overline{f^2}$
for an inaccurate weak meter does not have such simple form \cite{ME}.
\bibitem{FOOT} One can easily imagine a "bad"  question in a purely classical context.
Suppose there are two water drops on a surface. One puts a grain of sand into the first drop, 
and then tilts the surface so the two drops coalesce, with the grain of sand still inside.
In which of the two drops is it now? Under the new circumstances the question has no meaningful answer.
\bibitem{FOOT3} The argument is easily extended to a Hilbert space with dimension $N>2$, in which the measured quantity
$S$ is represented by an $N\times N$ matrix.
\end{thebibliography}
\end{document}